\documentclass[pre,twocolumn,superscriptaddress,floatfix]{revtex4-1}  

\usepackage[paperwidth=210mm,paperheight=297mm,centering,hmargin=2cm,vmargin=2.5cm]{geometry}

\usepackage[pdftex]{graphicx}
\usepackage{amsmath,amsfonts,amssymb}
\usepackage{natbib}
\usepackage{MnSymbol}
\usepackage{csquotes}
\usepackage{enumitem}

\usepackage{hyperref}
\usepackage{xcolor} 

\definecolor{bluemoi}{rgb}{0.25,0.50 ,0.75} 

\hypersetup{
  bookmarksopen=true,
  pdftitle=" ",
  pdfauthor=" ", 
  pdfsubject=" ", 
  pdftoolbar=true, 
  pdfmenubar=true, 
  pdfhighlight=/O, 
  colorlinks=true, 
  pdfpagemode=UseNone, 
  pdfpagelayout=SinglePage, 
  pdffitwindow=true, 
  linkcolor=bluemoi, 
  citecolor=bluemoi, 
  urlcolor=bluemoi 
}

\makeatletter
\renewcommand{\figurename}{\sf \textbf{Figure}}
\renewcommand{\thefigure}{\arabic{figure}}
\renewcommand{\fnum@figure}{\sf\textbf{\figurename}~\textbf{\thefigure}}
\renewcommand{\tablename}{\sf\textbf{Table}}
\renewcommand{\thetable}{\arabic{table}}
\renewcommand{\fnum@table}{\sf\textbf{\tablename}~\textbf{\thetable}}
\makeatother

\begin{document}

\title{Comprehensive decision-strategy space exploration\\ for efficient territorial planning strategies} 

\author{Olivier Billaud}
\thanks{Contributed equally to this work.}
\affiliation{TETIS, Univ Montpellier, AgroParisTech, Cirad, CNRS, INRAE, Montpellier, France}

\author{Maxence Soubeyrand}
\thanks{Contributed equally to this work.}
\affiliation{TETIS, Univ Montpellier, AgroParisTech, Cirad, CNRS, INRAE, Montpellier, France}

\author{Sandra Luque}
\affiliation{TETIS, Univ Montpellier, AgroParisTech, Cirad, CNRS, INRAE, Montpellier, France}

\author{Maxime Lenormand}
\thanks{Corresponding author maxime.lenormand@inrae.fr.}
\affiliation{TETIS, Univ Montpellier, AgroParisTech, Cirad, CNRS, INRAE, Montpellier, France}

\begin{abstract}
GIS-based Multi-Criteria Decision Analysis is a well-known decision support tool that can be used in a wide variety of contexts. It is particularly useful for territorial planning in situations where several actors with different, and sometimes contradictory, point of views have to take a decision regarding land use development. While the impact of the weights used to represent the relative importance of criteria has been widely studied in the recent literature, the impact of the order weights used to combine the criteria have rarely been investigated. This paper presents a spatial sensitivity analysis to assess the impact of order weights determination in GIS-based Multi-Criteria Analysis by Ordered Weighted Averaging. We propose a methodology based on an efficient exploration of the decision-strategy space defined by the level of risk and trade-off in the decision process. We illustrate our approach with a land use planning process in the South of France. The objective is to find suitable areas for urban development while preserving green areas and their associated ecosystem services. The ecosystem service approach has indeed the potential to widen the scope of traditional landscape-ecological planning by including ecosystem-based benefits, including social and economic benefits, green infrastructures and biophysical parameters in urban and territorial planning. We show that in this particular case the decision-strategy space can be divided into four clusters. Each of them is associated with a map summarizing the average spatial suitability distribution used to identify potential areas for urban development. We also demonstrate the pertinence of a spatial variance within-cluster analysis to disentangle the relationship between risk and trade-off values. At the end, we perform a site suitability ranking analysis to assess the relationship between the four detected clusters.
\end{abstract}

\maketitle

\section*{Introduction}

According to the World Health Organization, the fraction of population living in urban area has notably increased from the 35\% registered back in the 60's to the 55\% estimated in 2018; by 2050, 66\% of the world’s population is projected to be urban \citep{UN2019}. Intensive urban growth puts pressure on the resources and on the capacity of planners and local authorities to improve the standard of living of the residents. Sustainable development relies therefore on the successful management of urban sprawl and the development of efficient territorial planning strategies. 

This involves notably to identify the most suitable sites for locating future land uses based on explicit and/or implicit spatial information \citep{Collins2001}. The basic idea of site selection is to rank alternative sites based on their characteristics to identify the most suitable site for a specific land use \citep{Malczewski2004}. More specifically, GIS-based land-use suitability analysis consists in combining maps representing different suitability criteria to produce a suitability map showing the relative suitability of each site for a specific land use \citep{Klosterman2018}. GIS-based land-use suitability analysis has been applied in a wide variety of contexts, including but not limited to urban development. Examples of applications include the identification of animal habitats \citep{Store2001}, mapping of wilderness \citep{Comber2010}, ecosystem services assessment \citep{Langemeyer2016}, disease susceptibility mapping \citep{Dong2016,Tran2016}, soil fertility evaluation \citep{Mokarram2017} and urban planning \citep{Chen2014}, to name a few. 

Land use suitability analysis methods have evolved considerably since McHarg's seminal work in 1969 based on hand-drawn overlay techniques \citep{McHarg1969}. One of the more common GIS-based techniques used to assess land use suitability is the GIS-based Multi-Criteria Decision Analysis method (GIS-MCDA) \citep{Malczewski2015}. GIS-MCDA can be seen as \enquote{a process that transforms and combines geographical data and value judgments [...] to obtain appropriate and useful information for decision making} \citep{Malczewski2006,Tran2016}. It involves three main components: the geographical information standardization, the criterion weighting and the combination rules \citep{Malczewski2015}. The first component consists in standardizing the different geographical layers representing the criterion in comparable units and formats. The second component involves the weighting of criteria according to their importance. This weighting procedure is usually based on expert knowledge and depends on the structure of the problem. The most popular weighting methods are based on the concept of pairwise comparison matrix proposed by Saaty as part of Analytic Hierarchy/Network Process \citep{Saaty1987,Saaty2006}. The last component focuses on the combination of the different criteria and their associated weights. This combination essentially relies on additive weighting models. Of particular interests is the Ordered Weighted Averaging (OWA) operator \citep{Yager1988}. OWA introduces a second type of weights in the additive model (called order weights) that allows for the \enquote{modeling} of decision-maker's attitude in the decision through the fundamental concepts of risk, trade-off and decision-strategy space \citep{Jiang2000}.

As it is the case in any modeling framework, GIS-MCDA's outputs are subject to uncertainty generated at different levels of the process. Indeed, it exists several sources of uncertainty regarding GIS-MCDA's model, inputs and parameters that might impact the final suitability map on which the decision will be based. If we consider GIS-MCDA using OWA operator, four main sources of uncertainty can be identified in the aggregation process: the data used to build the criteria, the method used to generate the geographical layers (scale, standardization...), the methods used to determine the criteria weights and the order weights. While uncertainty of criteria weights has been already investigated in the recent literature \citep{Tran2016,Feizizadeh2017}, the sensitivity of GIS-MCDA outputs to order weight selection is a topic that is often neglected. The rigorous selection of order weight in GIS-MCDA is of major importance for discussing the outputs with policy makers \citep{Comber2010}. However, most of the studies focus on a few typical sets of order weights without developing clear method to efficiently explore the two-dimensional decision-strategy space. 

In this work, we tackle this problem by proposing an approach built around a method that allows for an automatic generation of order weights according to a given level of risk and trade-offs \citep{Lenormand2018}. The main advantage of this method lies in the formalization of the relationship between risk (likelihood that the decision made for a given location is wrong) and trade-off (degree of compensation between criteria) providing a clear definition of decision-strategy space. Thus, a full exploration of the decision-strategy space becomes possible, fulfilling a major gap regarding the sensitivity of GIS-MCDA to order weights selection. With an experimental design defined properly followed by a clustering analysis, the decision-strategy space can be segmented into several clusters of order weights leading to the production of similar suitability maps. The resulting clusters are further analyzed to extract an average suitability maps summarizing the information contains in each cluster. We can then perform a spatial variance within-cluster analysis to measure the degree of uncertainty associated with the average suitability maps. We finally investigate the relationship between clusters by performing a site suitability ranking analysis. In what follows, we will describe in detail the methodology before presenting the results highlighting the importance of order weights selection in practical applications.

We illustrate, here, our approach through an ecosystem services (ES) valuation for an urban development perspective in South of France, particularly concerned by the growing demands for urbanization. Allowing to evaluate a high diversity of benefits and impacts, the concept of ES has gained importance for urban and coastal planning \citep{Gomez2013, Arkema2015} and successfully combined with an GIS-MCDA framework \citep{Langemeyer2016} to undertake urban governance and related policies. Particularly, a promising use of ES as a decision-tool seems to reveal and anticipate trade-offs between the different ES and urbanization \citep{Langemeyer2016}. This is why we selected this concrete application to demonstrate the capacity of our approach for identifying potential spatial decisions regarding urban development in the light of different ES trade-offs.

The study area and the data are first introduced in the next section. We then describe in detail the methodology before presenting the results highlighting the importance of order weights selection in practical applications. The results and the potential limitations of our approach are finally discussed in the last section.

\begin{figure*}
	\begin{center}
		\includegraphics[width=13cm]{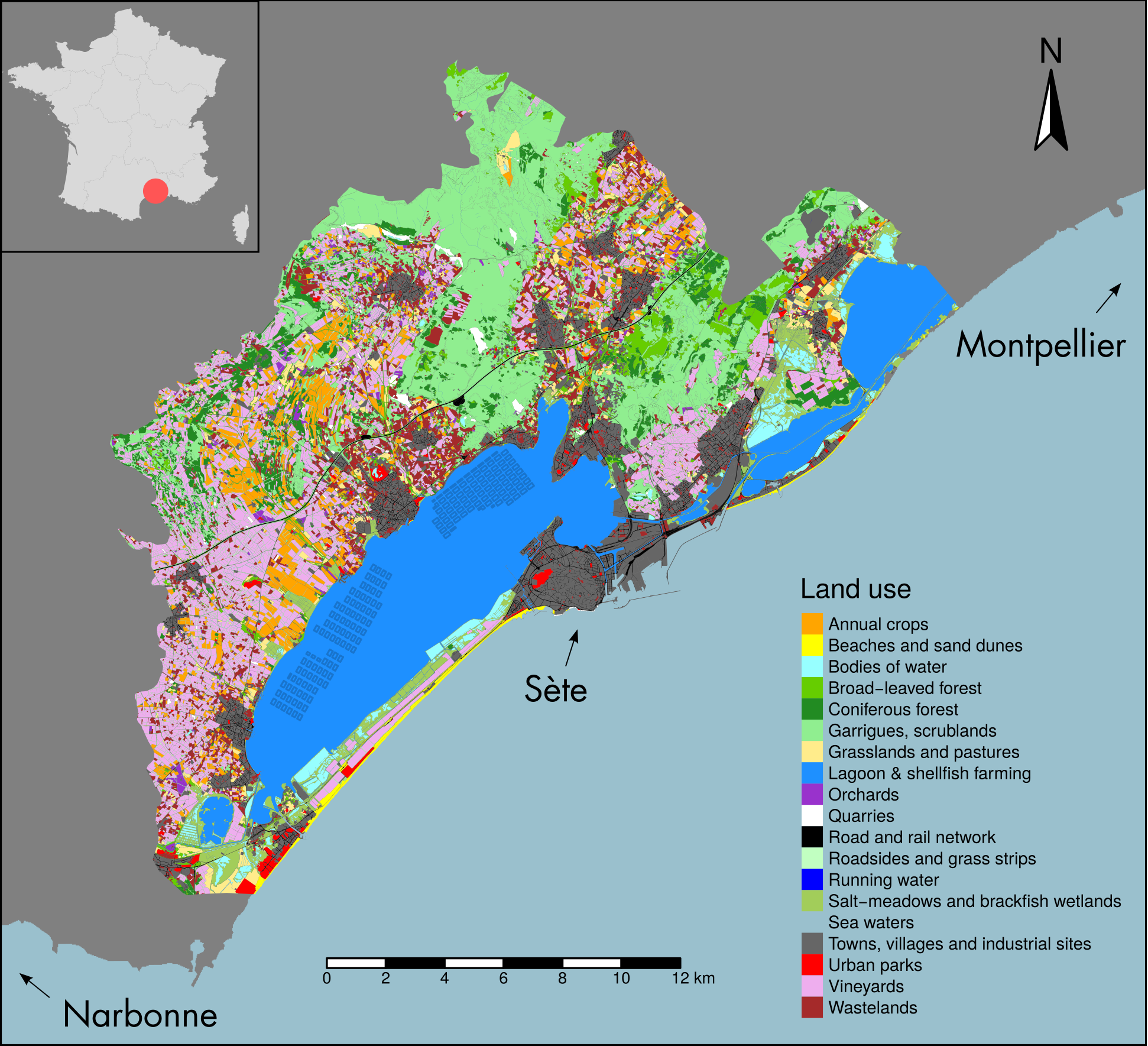}
		\caption{\sf \textbf{Land use map of Thau.} The inset shows the location of the studied areas (red circle) on a map of France. \label{Fig1}}
	\end{center}
\end{figure*} 

\section*{Study area, data and criteria}

The territory of Thau, situated on the French Mediterranean Coast about 20km West of Montpellier, is an important populated area of 60,000 ha and 117,000 inhabitants. With a central lagoon of 7500 ha and 35 km of coastlines, it is a land dominated by water. The territory is remarkably rich in terms of biodiversity, natural areas and landscapes, from the coastal and agricultural plains to the wetlands, the wooden reliefs and the  \emph{garrigue}. The region is highly attractive and subject to a heavy demographic pressure. To conciliate urban development, traditional economic activities such as wine production or shellfish farming and protection of the ecosystems is the main challenge of the territory. To do so, several instruments of water management and territorial planning exist and an engineering structure, the SMBT (Thau Bassin Joint Association, \emph{Syndicat Mixte du Bassin de Thau} in French), has been established. This contributed to build up a shared political vision and a common governance over the whole territory \citep{Maurel2014}. Our main challenge to support territorial planning needs in the region was to evaluate land suitability in order to identify the most suitable areas for spatial urban development. In agreement with the SMBT, which was our main partner for interacting with local stakeholders, we favored an approach based on the assessment of ecosystem services (ES) for prioritization of suitable areas susceptible of artificialization by urban development.

We evaluated a high number of ecosystem services based on local expert knowledge. We followed the methodological framework of the capacity matrix, developed by \citep{Burkhard2009} and improved by \citep{Jacobs2014} and \citep{Campagne2017}, which aims at assessing different land use and land cover  (LU/LC) types’ capacities to provide ecosystem services. In practice, we applied the methodological framework presented by \citep{Campagne2018}, using available LU/LC data (Figure \ref{Fig1}) and related biogeophysical information (see the Appendix for more details). In close collaboration with the SMBT, we determined a set of both, relevant services based on the Common International Classification of Ecosystem Services \citep{Haines2018} and LU/LC types drawn on an accurate land cover map over the territory \citep{Dupaquier2016}. Here we leverage expertise from a participatory workshop with local experts and contacted individuals supported by the SMBT to complete the matrix \citep{Campagne2017}. In order to assess the actual supply of ES for the study region, we applied the matrix method, as aforementioned, a qualitative valuation approach that links spatially explicit LU/LC types to a defined set of ES. Each expert proposes a score to assess the capacity of each LU/LC type (Figure \ref{Fig1}) to provide to each ES a value of importance. The capacity matrix method reduces the complexity of human-environmental systems, allowing reproducibility. It has proven useful in addressing the urgency-uncertainty dilemma for ES assessments, i.e., to provide best available knowledge where detailed modeling is not feasible or where data gaps obstruct more explicit approaches \citep{Jacobs2014}.

Nine ES were considered at the end, on the bases of the importance provided by the local experts and stakeholders from  the Thau region. These ES are listed in Table \ref{Criteria}. Two provisioning services focus on food from crop production and gathering. The four regulating services are related to main territorial challenges in terms of conservation of natural habitats for biodiversity (i.e. nesting sites, breeding grounds,...), including maintaining a good biochemical water status, protection against erosion and flooding. Cultural services also ranked high, reflecting the importance that people provide to the benefits from the surrounding nature, through contemplating beautiful landscape or open air activities. Finally, risks were also considered, in particular, the risk of wildfires that is pondered a main disservice of Mediterranean ecosystems.

\begin{table*}
	\begin{tabular}{lllc}
		Criterion & Type of service & Biophysical data & Criterion weight\\
		\hline
		Cultivated crops & Provisioning & Soil quality & 0.53\\
		Wild plants, algae, fungi and their outputs & Provisioning & - & 0.33\\
		Maintaining nursery populations and habitats & Regulating & ZNIEFF I and Natura 2000 areas & 1\\
		Maintenance of water quality & Regulating & Flow accumulation & 0.93\\
		Mass stabilization and control of erosion & Regulating & Slope Length Factor & 0.40\\
		Flood Protection & Regulating & Flooding hazard & 0.53\\
		Aesthetic & Cultural & - & 0.33\\
		Physical and experiential interactions & Cultural & Accessibility : distance to the main roads & 0.53\\
		Wildfires & Disservice & Fire hazard & 0.40\\
		Habitat patches importance & - & - & 1\\
		\hline
	\end{tabular}
	\caption{\textbf{List of criteria and their associated criterion weight.}}
	\label{Criteria}
\end{table*}

Each ES were spatialized using the high resolution ($5 \times 5 \mbox{m}^2$) LU/LC map (Figure \ref{Fig1}) and the average score given to the different LU/LC types by the experts. In order to refine the services assessment that is essentially based on the LU/LC type, we added external biophysical data to adjust the evaluation. We added for instance information about soil quality to the cultivated crops ES or a fire hazard index using physical informations (slope, exposure, wind, etc) to the disservice of wildfires. More details about this process are available in the Appendix. Note that we also added a tenth layer to measure the importance of habitat patches for landscape connectivity \citep{Saura2009} using the GUIDOS software \citep{Vogt2017}.

It is worth noting that the ten standardized geographical layers described above have been normalized to obtain a suitability for urban development inversely proportional to the score given by the expert. At the end of the process, for a given criteria, the pixel value are ranging from $0$ when it is unsuitable for urban land use to $1$ in the opposite case. All the criterion maps are available in Figure S1. All the information was stored in a matrix $Z$ which value $z_{ij}$ represents the suitability of pixel $i$ for urban land use according to criteria $j$.

Finally, the relative importance of each criteria (i.e. criterion weights) is the number of experts that considered the ES important for the territory of Thau divided by the total number of experts (Table \ref{Criteria}). An exception for the last criterion (the connectivity) that has been set to one after discussion with experts of the SMBT.

\begin{figure*}
	\begin{center}
		\includegraphics[width=14cm]{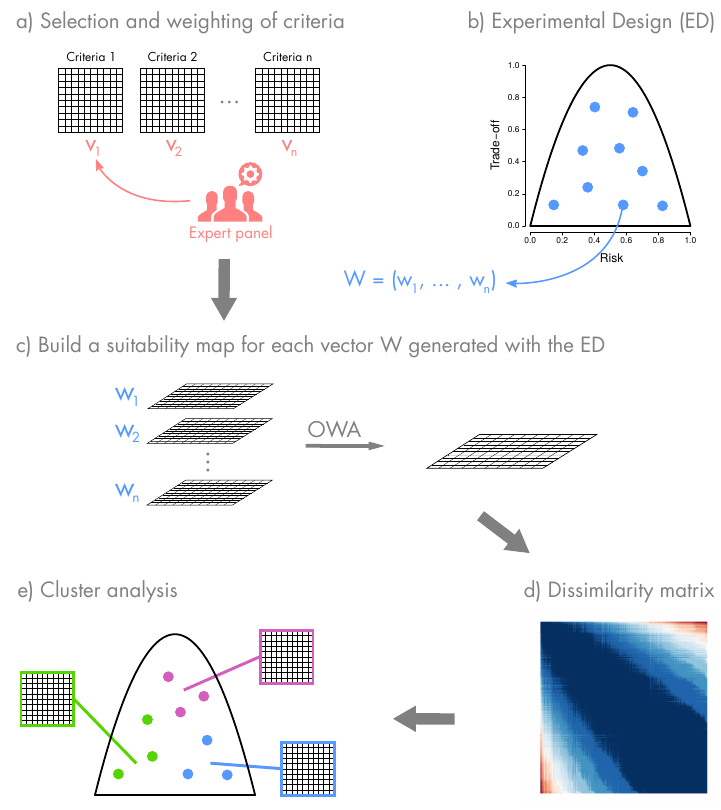}
		\caption{\sf \textbf{Methodology developed to efficiently explore the decision-strategy space.} a) Selection and weighting of criteria with an expert panel. b) Design an experimental plan to automatically generate order weights according to a set of risk and trade-off values. c) Build a suitability map for each vector of order weights generated with the experimental design. d) Build the dissimilarity matrix based on the euclidean distance between suitability map. e) Perform a cluster analysis to identify homogeneous area of the decision-strategy space. \label{Fig2}}
	\end{center}
\end{figure*}

\section*{Decision-strategy space's exploration}

This section describes the methods used to efficiently explore the decision-strategy space in GIS-MCDA using OWA operators. This methodology is summarized in Figure \ref{Fig2}.

\subsection*{OWA operator}

In this work, we rely on an Ordered Weighted Averaging (OWA) multi-criteria operator \citep{Yager1988} to combine all the criteria together in order to obtain a spatial explicit representation, that was a final map, to allow decision makers and local expertes to base planning decisions. OWA operators can indeed be used in a GIS context \citep{Jiang2000} in which every pixel is considered independently and defined by a set of criteria values. Two types of weights can be considered while using OWA operators: the \emph{criterion weights} representing the relative importance of the criteria in the decision process, and the \emph{order weights} characterizing the level of risk and trade-off taken in the decision. Criterion weights are represented by a vector $V=(v_j)_{1 \leq j \leq n}$, where $n$ stands for the number of criteria ($n=10$ in our case). The $j^{th}$ criterion weight corresponds to the relative importance given to the $j^{th}$ criteria. In this particular application, the criterion weights are given by a panel of experts (Table \ref{Criteria}) and reflects the importance of the criteria regarding urban land use. Order weights are represented by a vector $W=(w_j)_{1 \leq j \leq n}$. These weights are applied to the ranked criteria on a pixel basis. For a given pixel, the first order weight $w_1$ is assigned to the criterion with the lowest value, the second order weight $w_2$ to the second lowest criterion... While criterion weights are usually based on expert knowledge, order weights are used to adjust the level of risk and trade-off in the aggregation process. Note that in both cases, the vectors of weights sums to 1, $\sum_{j=1}^{n} v_j = \sum_{j=1}^{n} w_j =1$. Formally, the OWA operator is applied to every pixel $i$ with the following formula.
\begin{equation}
	OWA_i=\sum_{j=1}^{n} \left( \frac{v_{(j)} w_j}{\sum_{k=1}^{n} v_{(k)} w_k} \right) z_{i(j)} \label{OWA}
\end{equation}
where $z_{i(j)}$ is the $j^{th}$ lowest element of the collection of criteria $z_{ij}$ for the pixel $i$ and $v_{(j)}$ is the $j^{th}$ criterion weight reordered accordingly.

\subsection*{Risk and trade-off}

As mentioned above, order weights are deeply linked with the concepts of risk and trade-off in the decision-strategy process. Different order weights combinations $(w_1,...,w_n)\in [0,1]^n$ lead to different levels of risk and trade-off values, and we will see later that the reciprocal is not so clear. 

For a given pixel $i$, a combination of order weights that favors low $z_{ij}$ values, $W=(1,0,...,0)$ for example, represents a risk-aversion position. On the contrary, a combination of order weights that gives full weight to high $z_{ij}$ values, $W=(0,0,...,1)$ for example, represents a more risk-taking attitude. In the latter situation the risk is to give a high suitability to a pixel only based on the criteria with the highest suitability without taking into account the information given by the other criteria. As the name suggests, the trade-off represents the degree of accommodation between criteria. It can be seen as a measure of dispersion over the order weights. The maximum level of trade-off is reached when $w_j=1/n$ for all criteria and a minimum ($W=(1,0,...,0)$) and maximum ($W=(0,0,...,1)$) risk implies an absence of trade-off. Therefore, risk and trade-off are not independent and they are completely determined by the order weights distribution, the risk by its skewness and the trade-off by its dispersion \citep{Jiang2000}. 

The relationship between order weights, risk and trade-off forms a decision-strategy space usually represented by a triangle to highlight the inconsistency of certain couple of risk and trade-off values. If we express the level of risk and trade-off as a couple of values $(r,\,t) \in [0,1]^2$, the three vertices of the triangle are represented by the three main configurations: low risk with no trade-off $(r=0,\,t=0)$, high risk with no trade-off $(r=1,\,t=0)$ and medium risk with full trade-off  $(r=0.5,\,t=1)$.

\begin{figure*}
	\centering
	\includegraphics[width=12.25cm]{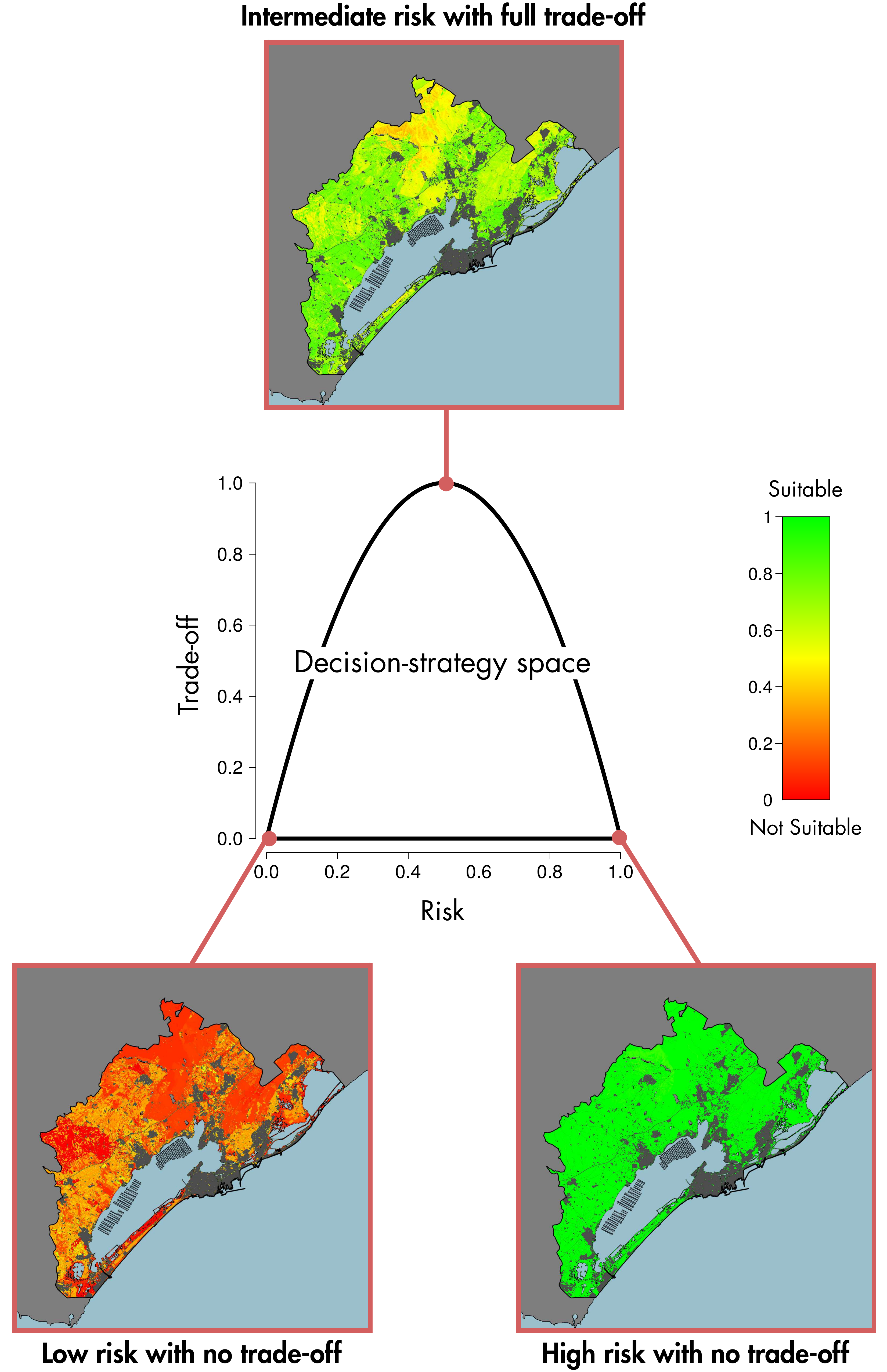}
	\caption{\sf \textbf{Decision-strategy space and suitability maps obtained with typical sets of order weights. \label{Fig3}}}
\end{figure*} 

\subsection*{Experimental design and cluster analysis}

Except for some trivial cases, like the ones described above, the limits of the decision-strategy space remain unclear, making an efficient exploration very complicated. This is due to the fact that, in practice, the generation of order weights according to a certain level of risk and trade-off is not formally established. \citep{Lenormand2018} recently proposed a methodology to generate order weights using truncated distributions. The method allows for an automatic generation of order weights according to a certain level of risk and trade-off based on the two first moments of the order weights distribution. More specifically, each suitable couple of $(r,\,t)$ values is associated with a truncated normal distribution with an average $r$ and a standard deviation proportional to $t$. The probability density function is then discretized to obtain $n$ order weights reflecting the predetermined level of risk $r$ and trade-off $t$. It is important to note that for certain risk and trade-off values, no suitable truncated normal distribution can be found. This allows for a formal delimitation of the decision-strategy space, which takes, in this case, the form of a parabola whose equation is $y=4x(1-x)$ (Figure \ref{Fig2}). See \citep{Lenormand2018} for more details.

In this work, we draw upon these recent advances in order weights determination to conduct a sensitivity analysis to assess the impact of the level of risk and trade-off in GIS-MCDA. The guiding idea is to efficiently explore the decision-strategy space in order to identify clusters of risk and trade-off values that leads to similar final suitability maps. As illustrated in Figure \ref{Fig2}, we develop on a five steps approach to reach this goal. As it is usually the case in GIS-MCDA, we start with the selection and weighting of criteria with a panel of experts selected for their knowledge and heterogeneous point of view regarding the problem (urban land use allocation in our case). We then design an experimental plan to automatically generate an order weight vector W covering the decision-strategy space. In this study, we generate $1,000$ vectors of order weights associated with $1,000$ risk and trade-off values. A suitability map is built for each vector of order weights using the OWA operator introduced in Equation \ref{OWA}. Then, the suitability maps are compared by measuring the euclidean distance between pixel values for each pair of suitability maps. We finally obtain a dissimilarity matrix used to perform a clustering analysis and identify clusters of risk and trade-off values. An average suitability map and its associated standard deviation is finally computed for each cluster.

\subsection*{Site suitability ranking analysis}

In order to compare the ranking of sites (i.e. $5 \times 5 \mbox{m}^2$ pixels) in two different clusters, we rely on the Kendall's $\tau$ coefficient \citep{Puka2011}. A value close to 1 means that the sites  are ordered in the same way according to their suitability in both clusters, while a value close to 0 means that there is no concordance in the cluster suitability rankings. It is also possible to evaluate the similarity between rankings by computing the percentage of sites in common in the top sites of the two rankings. The evolution of this quantity with the number of considered top sites will inform us on the relationship between the different detected clusters. 

\section*{Results}

\subsection*{Minimal decision-strategy space's exploration}

In order to get a better grasp of the situation we plot in Figure \ref{Fig3} suitability maps obtained with typical vectors of order weights. We observe that the low risk attitude ($W=(1,0,...,0)$) is very conservative and identify very few suitable areas (most of the pixel show a suitability lower than 0.5). The area identified as not suitable are mostly composed of natural vegetation. Nevertheless, this extreme position toward risk attitude allows for the identification of areas that are suitable to urban development whatever the considered criterion. There are indeed several pixels with a high suitable score (higher than 0.5) that are mainly close to built-up area. Conversely, an extreme risk taking attitude ($W=(0,0,...,1)$) tend to make the whole studied area suitable to urban development. This is due to the fact that for a given pixel one can always find a criterion suitable to urban development. These two configurations are not very nuanced since, on a pixel basis, the information is only coming from one criteria. Results obtained with the weighted linear combination that corresponds to an intermediate risk with full trade-off (top of the triangle) take better account of the heterogeneity of criteria. Note that in this particular configuration, the order weights have no impact on the process. We observe, then, that this full trade-off configuration leads globally to similar spatial features than the ones obtained in the low risk case, but with a level of suitability that is globally higher. However, beyond visual similarities between relative spatial suitability distributions, it is worth noting that the relative suitability of certain areas as compared with others is not the same in the two cases. Indeed, pixels with the same level of suitability in the low risk attitude case can have very different levels of suitability in the full trade-off configuration (see Figure S2 in Appendix for more details).

\subsection*{Comprehensive decision-strategy space's exploration}

\begin{figure}[h]
	\centering
	\includegraphics[width=\linewidth]{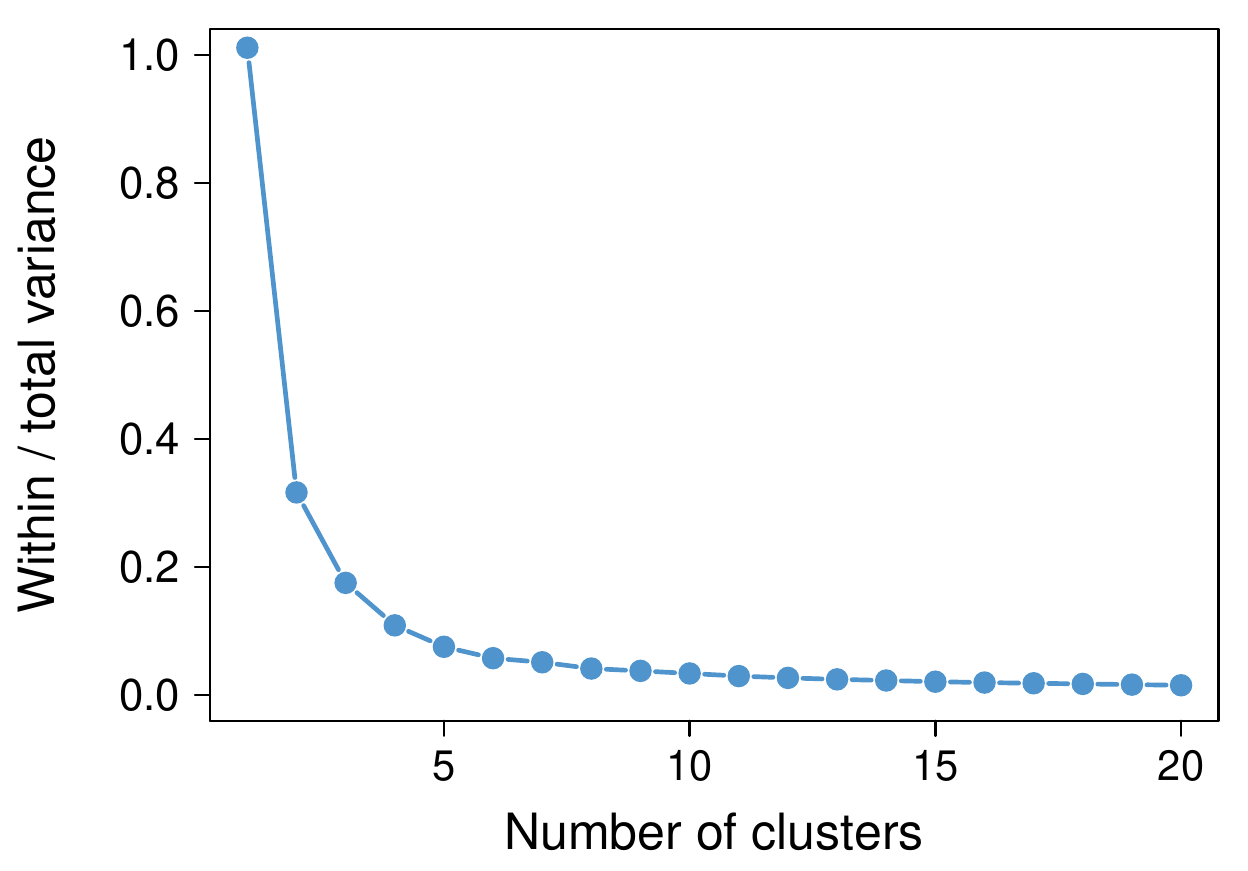}
	\caption{\sf \textbf{Ratio between the within-group variance and the total variance as a function of the number of clusters.\label{Fig4}}}
\end{figure}

This first result provides informative evidence on the importance of comparing the impact of risk and trade-off values on suitability maps in GIS-MCDA. This is why it is crucial to efficiently explore the decision-strategy space for a more accurate interpretation of the results obtained in GIS-MCDA. Following the method described in the previous section, we build a dissimilarity matrix between $1,000$ suitability maps associated with $1,000$ couple of risk and trade-off values drawn at random within the parabolic decision-strategy space. In order to identify potential cluster of risk and trade-off values leading to similar suitability maps, we apply an ascending hierarchical clustering (AHC) algorithm on this dissimilarity matrix using the Ward's agglomeration method. We choose the number of cluster based on the evolution of the ratio between the within-group variance and the total variance as a function of the number of clusters (Figure \ref{Fig4}).

\begin{figure*}
	\centering
	\includegraphics[width=12.4cm]{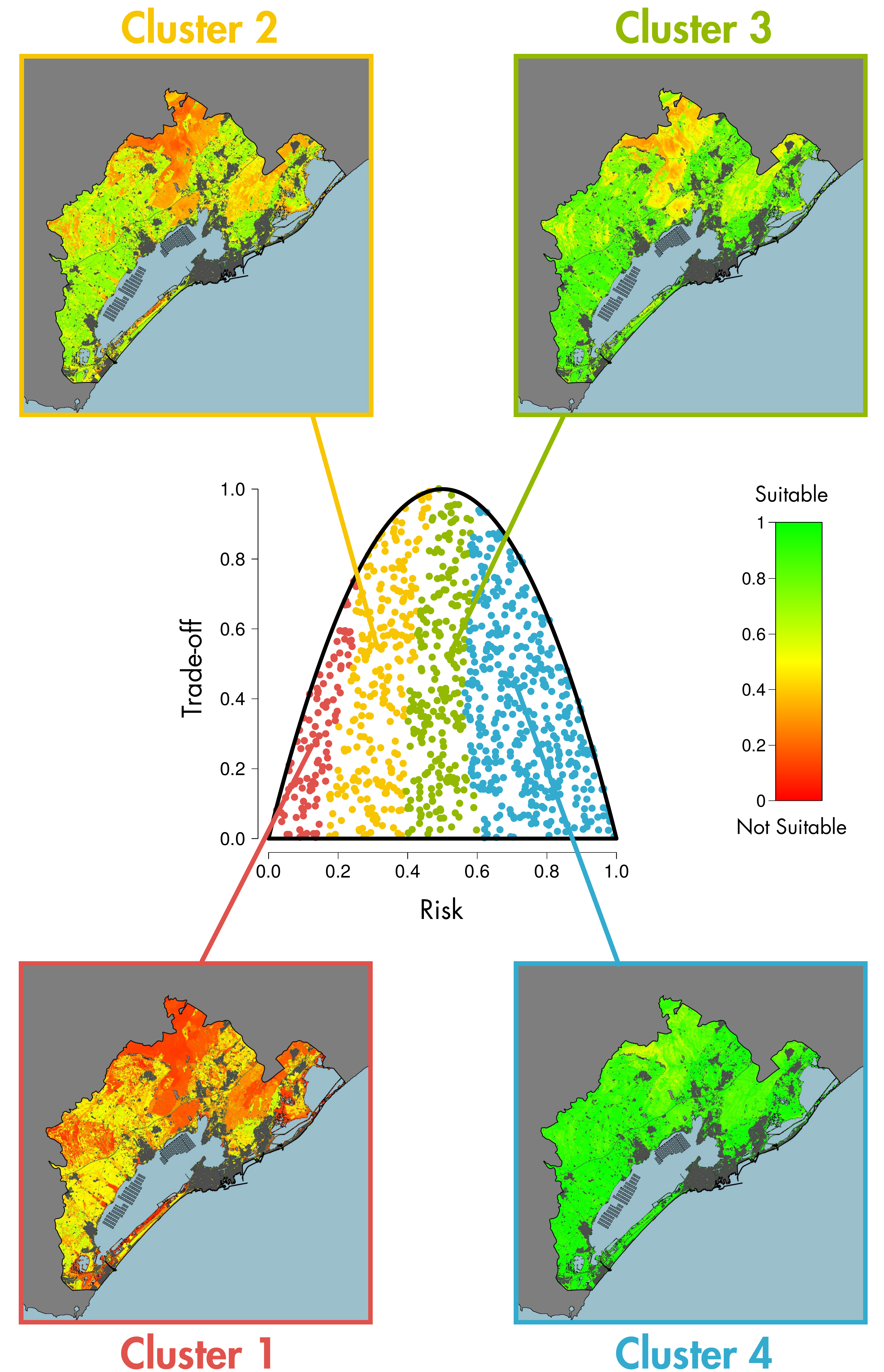}
	\caption{\sf \textbf{Average suitability maps associated with the four clusters. \label{Fig5}}}
\end{figure*}

\begin{figure*}
	\centering
	\includegraphics[width=12.4cm]{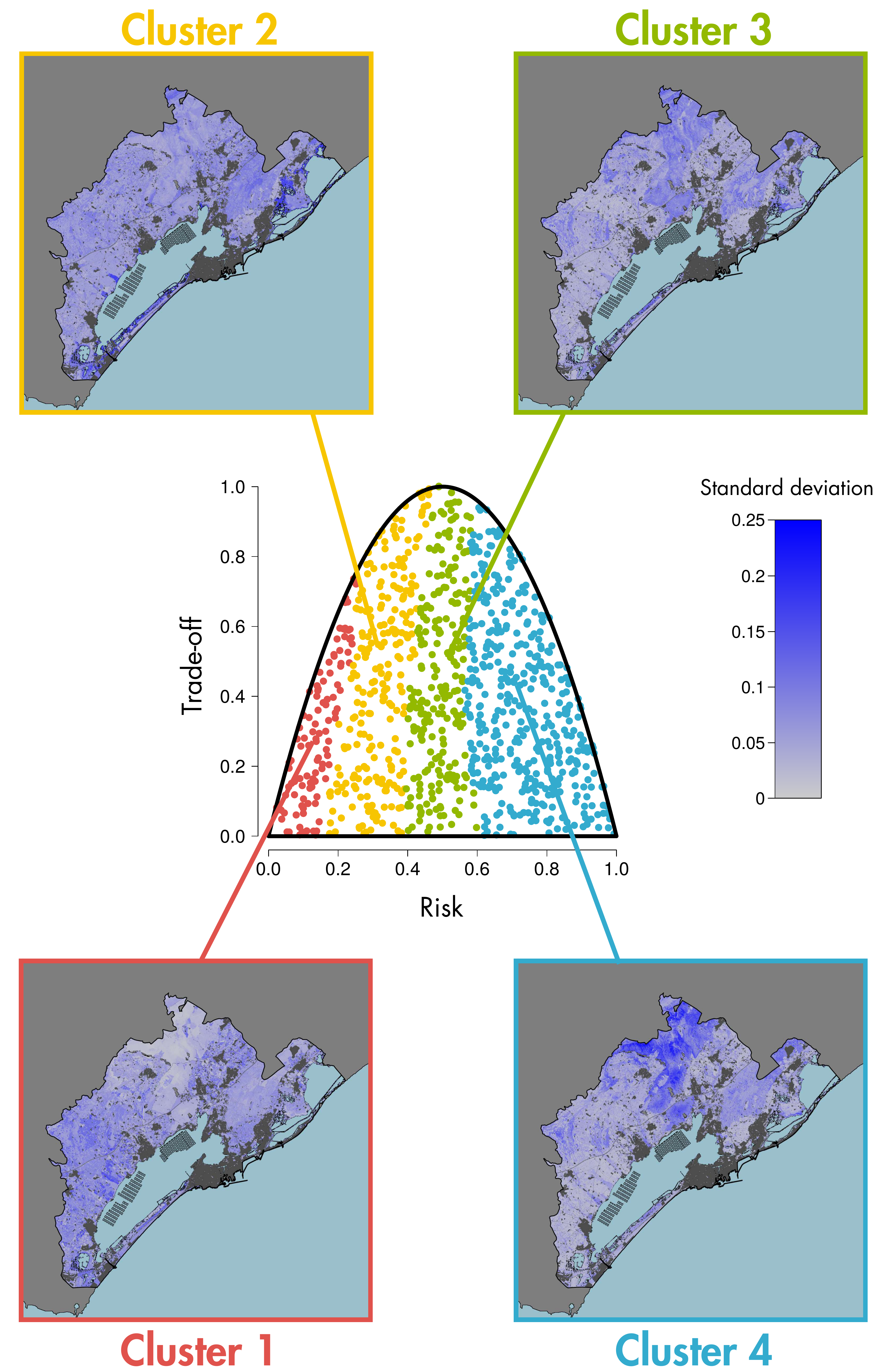}
	\caption{\sf \textbf{Spatial distribution of the within-cluster variability. \label{Fig6}}}
\end{figure*}

We detect four clusters or groups of risk and trade-offs values that share similarities in the suitability maps that they produced. We then compute the average suitability maps associated with each of the selected clusters. The results are shown in Figure \ref{Fig5}. We first observe that the segmentation of the parabolic decision-strategy space is mainly driven by the level of risk. Although there is a difference in trend across pixels, the global level of suitability increases monotonously with the risk. The segmentation obtained with high levels of hierarchical aggregation reflects therefore the global structure induced by the reordering of criteria. However, it must be noticed that when considering a higher number of clusters a trade-off based segmentation highlights differences among pixels for a given global level of suitability (see Figure S3 in Appendix for more details). The average suitability maps associated with the four clusters enable us to have a more nuanced view of the situation in the Thau region that with the typical sets of order weights (Figure \ref{Fig3}). The low risk cluster provided a very conservative situation but bring into light the most suitable areas (the only ones with a correct suitability) that have a potential to be urbanized first within a sustainable planning framework. The low-medium risk cluster releases a large proportion of the agricultural land-use that is still suitable for urbanization. Conversely it allowed to identify the most important agricultural areas. The map of the medium-high risk cluster identifies almost all the agricultural areas suitable for urbanization. Ecosystems that provide a maximum of ES such as forests, garrigues or salt-meadows are the only areas that are considered not suitable using the approach presented. Eventually, the high risk cluster provided a map quite similar to the extreme risk situation, but allows to detect some critical ecosystems (mostly forests) of major importance in the region.

Although the total within-group variance obtained with four clusters is relatively low (Figure \ref{Fig4}), it is also important to evaluate how this variability is spatially distributed. Notably, this variance based analysis will allow us to assess the impact of the trade-off for a given level of risk. We observe in Figure \ref{Fig6} that the spatial distribution of standard deviation of the suitability varies from one cluster to another. The uncertainty seems to be correlated with the suitability. However, the nature of the relationship changes according to the type of cluster. The uncertainty tends to increase with the suitability for the low and medium risk cluster, while the opposite behavior is observed for the high risk cluster. 

\subsection*{Site suitability ranking analysis}

Table \ref{Tau} shows the Kendall's $\tau$ coefficient matrix \citep{Puka2011} between the site suitability rankings obtained with the different clusters (Figure \ref{Fig5}). We observe that the coefficient are globally high with values ranging from 0.6 and 0.85. We find in particular that $\tau$ decreases as the distance between clusters increases. 

\begin{table}[ht]
	\caption{\textbf{Kendall rank correlation matrix.} Kendall's $\tau$ coefficient between the four suitability maps displayed in Figure \ref{Fig5}. Scatterplots of the suitability values obtained with the different clusters are available in Figure S4 in Appendix.}
	\begin{center}
		\begin{tabular}{ccccc}
			& Cluster 1 & Cluster 2 & Cluster 3 & Cluster 4\\ 
			Cluster 1 & 1 & 0.84 & 0.72 & 0.61\\ 
			Cluster 2 &  & 1 & 0.85 & 0.71\\
			Cluster 3 &  &  & 1 & 0.82\\ 
		\end{tabular}
	\end{center}
	\label{Tau}
\end{table}

\begin{figure}
	\centering
	\includegraphics[width=\linewidth]{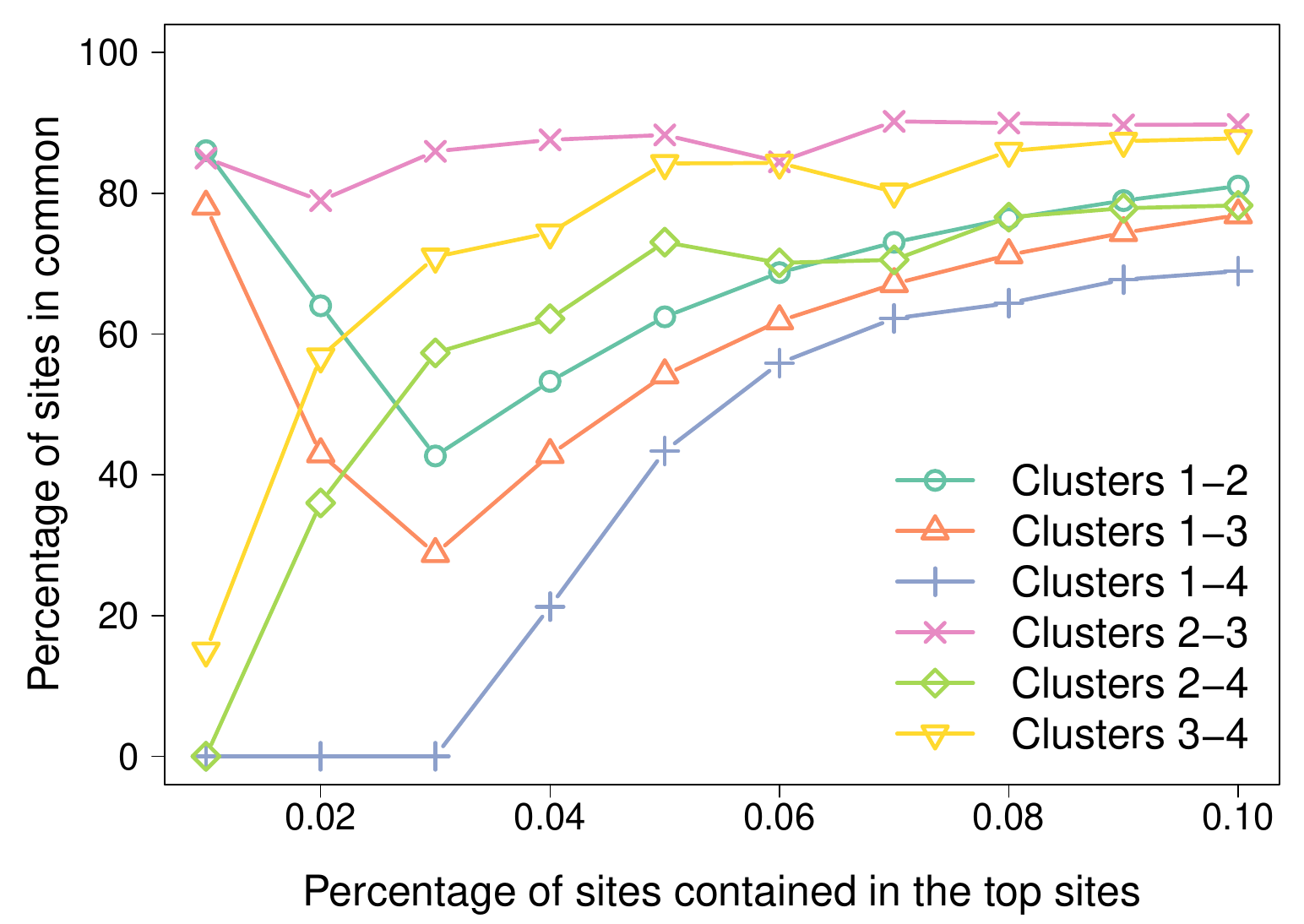}
	\caption{\sf \textbf{Pairwise cluster ranking comparisons.} For each pair of clusters the percentage of sites in common in the top suitable sites is computed. For example, there is about 80\% of sites (i.e. pixels) in common in the top 0.01\% of suitable sites obtained with the clusters 1 and 3 (first red triangle). \label{Fig7}}
\end{figure}

A similar pattern can be observed in Figure \ref{Fig7}. The percentage of sites in common in the top sites of two rankings tends to decrease with the distance between clusters. In most cases, the similarity between clusters increases with the number of sites considered in the top sites. Clusters 2 and 3 are very similar in terms of rankings with more than 80\% of sites in common whatever the top sites considered. The similarity between the cluster 1 and clusters 2 and 3 shows a different pattern with a strong similarity of sites in the top 0.01\% followed by a rapid decay and then an increase as observed in the general trend. This implies that clusters 1, 2 and 3 share a core of high suitability sites. Although, cluster 4 does not share this core of high suitability sites, it is interesting to note that, considering the top 0.03\% and higher, it shares more sites with cluster 2 and 3 than cluster 1 does.

\section*{Discussion}

In this paper, we presented an approach to assess the impact of order weights and their associated level of risk and trade-off in GIS-based Multi-Criteria Decision Analysis. The developed approach aims at efficiently explore the decision-strategy space formed by the relationship between risk and trade-off in Ordered Weighted Averaging. Although the results might vary according to the formula used, a vector of order weights in OWA, whatever its size, can be summarized by its level of risk and trade-off. However, the opposite is not always true, making it complicated to automatically generate order weights according to a certain level of risk and trade-off and conduct rigorous sensitivity analysis. In this paper, we relied on the recent advances made in order weight determination proposed in \citep{Lenormand2018}. The method presented in this paper proposes to automatically generate order weights using truncated distributions according to a certain level of risk and trade-off based on the two first moments of the order weights distribution. This methodological framework enabled us to generate a suitability map for each couple of suitable risk and trade-off values through an OWA operator, opening the possibility to perform systematic sensitivity and uncertainty analysis in GIS-MCDA. It is indeed important to model the decision makers' attitudes toward risk in MCDA \citep{Bodily1985,Malczewski2015}. The comprehensive exploration of the two-dimensional decision-strategy space proposed in this paper provides a generic and rigorous way to model, aggregate and compare a full range of scenarios. This is an important topic in MCDA that should not be neglected. Especially since it it can be easily combined with existing method used to compute the criteria weights \citep{Eldrandaly2013,Mokarram2018,Kazemi2019,Zabihi2019}.

We illustrated the potential of our approach on a case study of land use management in the territory of Thau (France). We selected this example to propose practical solutions to the common problem that is experiencing the south of France regarding urban sprawl, but also for its relevance as a concrete application of ecosystems' valuations using GIS-MCDA. The idea here is to prioritize suitable areas for urban development in the Thau territory, while preserving optimally \enquote{green} areas and their associated ecosystem services to promote a sustainable regional management. Indeed, one of the biggest challenges we are facing nowadays is to find suitable new areas for development in a sustainable way, without degrading existing valuable land uses. Spatial planning for urban development is directly related to decision-making and government strategies, besides outcomes of planning will clearly differ depending on stakeholders and territorial policies. The ecosystem service approach has the potential to widen the scope of traditional landscape-ecological planning to include ecosystem-based benefits, including social and economic benefits, green infrastructures and biophysical parameters to directly strengthen the role of landscape-ecological planning in urban and territorial planning \citep{Kopperoinen2014,Bezak2017}. The results obtained in this study demonstrates that a complete exploration of the decision-strategy space enabled us to get a better understanding of how the different ecosystem services combine to produce the final suitability map. In particular, the clustering analysis led to an informative segmentation of the decision-strategy space that helped us to really understand the role of order weights and particularly their impact on the trade-off between criteria. Finally, a site suitability ranking analysis allowed us to examine the differences and similarities between the different clusters.

Our approach has the great advantage to be generic and relatively easy to implement in any multi-criteria evaluations. Although further analysis on more case studies are certainly needed, we believe that the approach developed in this study could provide an operational and collaborative tool to facilitate decision making processes to improve land use planning. This is particularly important as the final suitability map obtained in such process might be used to take important and irreversible decisions regarding land use management. Therefore, it is crucial to provide robust and quality-assessed materials to facilitate the dialogue between stakeholders and decision-makers.

\section*{Acknowledgments}

This study was partially supported by the ERANET BIODIVERSA 3 funded Project IMAGINE (Integrative Management of Green Infrastructures Multifunctionality, Ecosystem integrity and Ecosystem Services: From assessment to regulation in socio-ecological systems) and the French National Research Agency (project NetCost, ANR-17-CE03-0003 grant). A special thanks goes to Pierre Maurel, Julien Le Viol and the members of the SMBT of Thau Region. We acknowledge Marine Le Louarn for providing us with the landscape connectivity map. 

\section*{Data availability}

Code and data are available at \url{www.maximelenormand.com/Codes}

\vspace*{0.5cm}
\section*{Author contributions}

ML coordinated the study, designed the study and wrote the paper. ML, OB and MS wrote the code, processed and analyzed the data. SL coordinated the study. All authors read, commented and validated the final version of the manuscript.

\bibliographystyle{unsrt}
\bibliography{MCDA}

\begin{thebibliography}{10}

\bibitem{UN2019}
{United Nations, Department of Economic and Social Affairs, Population
  Division}.
\newblock {World Population Prospects 2019 Highlights (ST/ESA/SER.A/423)},
  2019.

\bibitem{Collins2001}
M.~G. Collins, F.~R. Steiner, and M.~J. Rushman.
\newblock Land-{{Use Suitability Analysis}} in the {{United States}}:
  {{Historical Development}} and {{Promising Technological Achievements}}.
\newblock 28(5):611--621, 2001.

\bibitem{Malczewski2004}
J.~Malczewski.
\newblock {{GIS}}-based land-use suitability analysis: A critical overview.
\newblock {\em Progress in Planning}, 62(1):3--65, 2004.

\bibitem{Klosterman2018}
R.~E. Klosterman, K.~Brooks, J.~Drucker, E.~Feser, and H.~Renski.
\newblock {\em Planning {{Support Methods}}: {{Urban}} and {{Regional
  Analysis}} and {{Projection}}}.
\newblock {Rowman \& Littlefield}, June 2018.

\bibitem{Store2001}
R.~Store and J.~Kangas.
\newblock Integrating spatial multi-criteria evaluation and expert knowledge
  for {{GIS}}-based habitat suitability modelling.
\newblock {\em Landscape and Urban Planning}, 55(2):79--93, 2001.

\bibitem{Comber2010}
A.~Comber, S.~Carver, S.~Fritz, R.~McMorran, J.~Washtell, and P.~Fisher.
\newblock Different methods, different wilds: {{Evaluating}} alternative
  mappings of wildness using fuzzy {{MCE}} and {{Dempster}}-{{Shafer MCE}}.
\newblock {\em Computers, Environment and Urban Systems}, 34(2):142--152, 2010.

\bibitem{Langemeyer2016}
J.~Langemeyer, E.~G{\'o}mez-Baggethun, D.~Haase, S.~Scheuer, and T.s Elmqvist.
\newblock Bridging the gap between ecosystem service assessments and land-use
  planning through {Multi}-{Criteria} {Decision} {Analysis} ({MCDA}).
\newblock {\em Environmental Science \& Policy}, 62:45--56, 2016.

\bibitem{Dong2016}
S.~Dong.
\newblock Comparisons between different multi-criteria decision analysis
  techniques for disease susceptibility mapping.
\newblock {\em Student thesis series INES}, 2016.

\bibitem{Tran2016}
A.~Tran, C.~Trevennec, J.~Lutwama, J.~Sserugga, M.~G{\'e}ly, C.~Pittiglio,
  J.~Pinto, and V.~Chevalier.
\newblock Development and {Assessment} of a {Geographic} {Knowledge}-{Based}
  {Model} for {Mapping} {Suitable} {Areas} for {Rift} {Valley} {Fever}
  {Transmission} in {Eastern} {Africa}.
\newblock {\em PLOS Neglected Tropical Diseases}, 10(9):e0004999, 2016.

\bibitem{Mokarram2017}
M.~Mokarram and M.~Hojati.
\newblock Using ordered weight averaging ({{OWA}}) aggregation for
  multi-criteria soil fertility evaluation by {{GIS}} (case study: Southeast
  {{Iran}}).
\newblock {\em Computers and Electronics in Agriculture}, 132, 2017.

\bibitem{Chen2014}
J.~Chen.
\newblock {{GIS}}-based multi-criteria analysis for land use suitability
  assessment in {{City}} of {{Regina}}.
\newblock {\em Environmental Systems Research}, 3(1):13, 2014.

\bibitem{McHarg1969}
I.~L. McHarg.
\newblock {\em Design with {{Nature}}}.
\newblock {John Wiley \& Sons}, 1969.

\bibitem{Malczewski2015}
J.~Malczewski and C.~Rinner.
\newblock {\em Multicriteria {{Decision Analysis}} in {{Geographic Information
  Science}}}.
\newblock Advances in {{Geographic Information Science}}. {Springer-Verlag},
  {Berlin Heidelberg}, 2015.

\bibitem{Malczewski2006}
J.~Malczewski.
\newblock {{GIS}}-based multicriteria decision analysis: A survey of the
  literature.
\newblock {\em International Journal of Geographical Information Science},
  20(7):703--726, 2006.

\bibitem{Saaty1987}
R.~W. Saaty.
\newblock The analytic hierarchy process\textemdash{}what it is and how it is
  used.
\newblock {\em Mathematical Modelling}, 9(3):161--176, 1987.

\bibitem{Saaty2006}
T.~L. Saaty.
\newblock The {{Analytic Network Process}}.
\newblock In T.~L. Saaty and L.~G. Vargas, editors, {\em Decision {{Making}}
  with the {{Analytic Network Process}}: {{Economic}}, {{Political}},
  {{Social}} and {{Technological Applications}} with {{Benefits}},
  {{Opportunities}}, {{Costs}} and {{Risks}}}, International {{Series}} in
  {{Operations Research}} \& {{Management Science}}, pages 1--26. {Springer
  US}, 2006.

\bibitem{Yager1988}
R.~R. Yager.
\newblock {On Ordered Weighted Averaging Aggregation Operators in Multicriteria
  Decision making}.
\newblock {\em IEEE Transactions on Systems, Man, and Cybernetics},
  18(1):183--190, 1988.

\bibitem{Jiang2000}
H.~Jiang and J.~R. Eastman.
\newblock {Application of Fuzzy Measures in Multi-Criteria Evaluation in GIS}.
\newblock {\em International Journal of Geographical Information Science},
  14(2):173--184, 2000.

\bibitem{Feizizadeh2017}
B.~Feizizadeh and S.~Kienberger.
\newblock Spatially explicit sensitivity and uncertainty analysis for
  multicriteria-based vulnerability assessment.
\newblock {\em Journal of Environmental Planning and Management},
  60(11):2013--2035, 2017.

\bibitem{Lenormand2018}
M.~Lenormand.
\newblock Generating {OWA} weights using truncated distributions.
\newblock {\em International Journal of Intelligent Systems}, 33(4):791--801,
  2018.

\bibitem{Gomez2013}
E.~{G{\'o}mez-Baggethun} and D.~N. Barton.
\newblock Classifying and valuing ecosystem services for urban planning.
\newblock {\em Ecological Economics}, 86(C), 2013.

\bibitem{Arkema2015}
K.~K. Arkema, G.~M. Verutes, S.~A. Wood, C.~{Clarke-Samuels}, S.~Rosado, M.a
  Canto, A.~Rosenthal, M.~Ruckelshaus, G.~Guannel, J.~Toft, J.~Faries, J.~M.
  Silver, R.~Griffin, and A.~D. Guerry.
\newblock Embedding ecosystem services in coastal planning leads to better
  outcomes for people and nature.
\newblock {\em Proceedings of the National Academy of Sciences},
  112(24):7390--7395, 2015.

\bibitem{Maurel2014}
P.~Maurel, R.~Plant, O.~Barreteau, and Y.~Bertacchini.
\newblock Beyond iwrm : developing territorial intelligence in the thau
  territory, france.
\newblock In Victor Squires, Hugh Milner, and Katherine Daniell, editors, {\em
  River Basin Management in the Twenty-First Century : People and Place}, pages
  22--41. CRC Press, 2014.

\bibitem{Burkhard2009}
B.~Burkhard, F.~Kroll, F.~M{\"u}ller, and W.~Windhorst.
\newblock Landscapes' capacities to provide ecosystem services ? a concept for
  land-cover based assessments.
\newblock {\em Landscape Online}, 15:1--12, 2009.

\bibitem{Jacobs2014}
S.~Jacobs, B.~Burkhard, T.~Daele, J.~Staes, and A.~Schneiders.
\newblock 'the matrix reloaded': A review of expert knowledge use for mapping
  ecosystem services.
\newblock {\em Ecological Modelling}, pages 21--30, 2014.

\bibitem{Campagne2017}
C.~S. Campagne, P.~Roche, F.~Gosselin, L.~Tschanz, and T.~Tatoni.
\newblock Expert-based ecosystem services capacity matrices: Dealing with
  scoring variability.
\newblock {\em Ecological Indicators}, 79:63--72, 2017.

\bibitem{Campagne2018}
C.~S. Campagne and P.~Roche.
\newblock May the matrix be with you! guidelines for the application of
  expert-based matrix approach for ecosystem services assessment and mapping.
\newblock {\em One Ecosystem}, One Ecosystem 3: e24134, 05 2018.

\bibitem{Haines2018}
R~Haines-Young and M.B Potschin.
\newblock Common international classification of ecosystem services (cices)
  v5.1 and guidance on the application of the revised structure, 2018.

\bibitem{Dupaquier2016}
C.~Dupaquier, A.~Desbrosse, P.~Maurel, and J-P Roussillon.
\newblock Cartographie de l'occupation du sol sur le bassin de {T}hau.
  {R}apport m{\'e}thodologique, 2016.

\bibitem{Saura2009}
S.~Saura and Josep Torn{\'e}.
\newblock Conefor {Sensinond}e 2.2: A software package for quantifying the
  importance of habita patches for landscape connectivity.
\newblock {\em Environmental Modelling \& Software}, 24:135--139, 2009.

\bibitem{Vogt2017}
P.~Vogt and K.~Riitters.
\newblock {GuidosToolbox}: universal digital image object analysis.
\newblock {\em European Journal of Remote Sensing}, 50:352--361, 2017.

\bibitem{Puka2011}
L.~Puka.
\newblock {\em Kendall's Tau}, pages 713--715.
\newblock Springer Berlin Heidelberg, Berlin, Heidelberg, 2011.

\bibitem{Bodily1985}
S.~E. Bodily.
\newblock {\em Modern {{Decision Making}}: {{A Guide}} to {{Modeling With
  Decision Support Systems}}}.
\newblock {McGraw Hill Higher Education}, {New York}, 1985.

\bibitem{Eldrandaly2013}
K.~A. Eldrandaly.
\newblock Exploring multi-criteria decision strategies in {{GIS}} with
  linguistic quantifiers: An extension of the analytical network process using
  ordered weighted averaging operators.
\newblock {\em International Journal of Geographical Information Science},
  27(12):2455--2482, 2013.

\bibitem{Mokarram2018}
Marzieh Mokarram and Abbas Mirsoleimani.
\newblock Using {{Fuzzy}}-{{AHP}} and order weight average ({{OWA}}) methods
  for land suitability determination for citrus cultivation in {{ArcGIS}}
  ({{Case}} study: {{Fars}} province, {{Iran}}).
\newblock {\em Physica A: Statistical Mechanics and its Applications},
  508:506--518, October 2018.

\bibitem{Kazemi2019}
M.~{Kazemi-Beydokhti}, R.~Ali Abbaspour, M.~Kheradmandi, and
  A.~{Bozorgi-Amiri}.
\newblock Determination of the physical domain for air quality monitoring
  stations using the {{ANP}}-{{OWA}} method in {{GIS}}.
\newblock {\em Environmental Monitoring and Assessment}, 191(2):299, 2019.

\bibitem{Zabihi2019}
H.~Zabihi, Me~Alizadeh, Pe~Kibet~Langat, Me~Karami, He~Shahabi, Ae~Ahmad,
  M.~Nor~Said, and S.~Lee.
\newblock {{GIS Multi}}-{{Criteria Analysis}} by {{Ordered Weighted Averaging}}
  ({{OWA}}): {{Toward}} an {{Integrated Citrus Management Strategy}}.
\newblock {\em Sustainability}, 11(4):1009, 2019.

\bibitem{Kopperoinen2014}
L.~Kopperoinen, P.~Itkonen, and J.~Niemel{\"a}.
\newblock Using expert knowledge in combining green infrastructure and
  ecosystem services in land use planning: An insight into a new place-based
  methodology.
\newblock 29(8):1361--1375, 2014.

\bibitem{Bezak2017}
P.~Bez{\'a}k, P.~Mederly, Z.~Izakovi{\v c}ov{\'a}, J.~{\v S}pulerov{\'a}, and
  C.~Schleyer.
\newblock Divergence and conflicts in landscape planning across spatial scales
  in {{Slovakia}}: {{An}} opportunity for an ecosystem services-based approach?
\newblock {\em International Journal of Biodiversity Science, Ecosystem
  Services \& Management}, 13(2):119--135, 2017.

\end{thebibliography}

\onecolumngrid
\onecolumngrid

\makeatletter
\renewcommand{\fnum@figure}{\sf\textbf{\figurename~\textbf{S}\textbf{\thefigure}}}
\renewcommand{\fnum@table}{\sf\textbf{\tablename~\textbf{S}\textbf{\thetable}}}
\makeatother

\setcounter{figure}{0}
\setcounter{table}{0}
\setcounter{equation}{0}

\section*{Appendix}

\subsection*{Biophysical data}

As mention in the main text, we added external biophysical data (Table \ref{Criteria}) to adjust the ES valuation. For each pixel, the experts’ scores associated with the ES has been multiply by the factor extracted from the biophysical data as described below. 

\begin{itemize}
	\item The soil quality is an index made by the French National Institute for Agronomic Research and based on two main constraints (salinity and the available water capacity) and secondary constraints (soil sealing, hydromorphy, stoniness and pH). It has 16 categories so we assigned a multiplicative factor 1 to the most fertile and decreased of 0.05 for each category to end at 0.25 for salted ground. 
	\item Habitats of ecological importance are defined areas by naturalists’ studies such as ZNIEFF I (Natural zone of ecological interest in terms of fauna and flora) or Natura 2000 (EU wide network of nature protection areas). We applied factor 1 when the pixel was located inside these protected area and 0.75 otherwise.
   	\item Flow accumulation follow a hydrologic model of Multiple Flow Direction Algorithm. The value are converted to a 0-1 continuous scale. To avoid a crushed scale, we used 98\% of the maximal value to set 1. Values above are also set to 1. 
	\item We used a Slope Length Factor used by the Universal Soil Loss Equation (USLE) to adjust the mass stabilization and control of erosion ES. Here also, the value are converted to a 0-1 continuous scale. To avoid a crushed scale, we used 98\% of the maximal value to set 1. Values above are also set to 1. 
	\item The flooding hazard came from the official and mandatory risk prevention plans for inundations. It has three categories (high, medium, none) that corresponds respectively to a factor 1, 0.5 or 0. 
	\item Distance to the main road was computed on SAGA GIS, taking into account of the relief. It is a proxy of accessibility, we used the threshold of 300m as the maximum distance for a recreational walk (5 min walk) to defined three categories: $\leq$ 300m, 300m-1km, $\geq$ 1km. Since almost the whole territory is suitable for hiking (no high elevation and slopes) we stopped the scale at 0.5 for a distance higher than 1km (around 15min walk) and computed a decreasing and continuous scale from 1 to 0.5 between 300m and 1km.
	\item The fire hazard has been derived from forest stands combustibility and spreading factors such as slope and speed propagation in a study carried out by the National Forest Office (ONF) and the Sea and Territory Departmental Direction (DDTM). For 6 values (very high, high, medium, low, very low, none) we started at a factor 1 for a very high risk and decreased of 0.1 for each category to end at 0.5 (unlike the flooding hazard, we considered that there was always a fire risk).
\end{itemize}

\begin{figure*}[!h]
	\begin{center}
		\includegraphics[width=13cm]{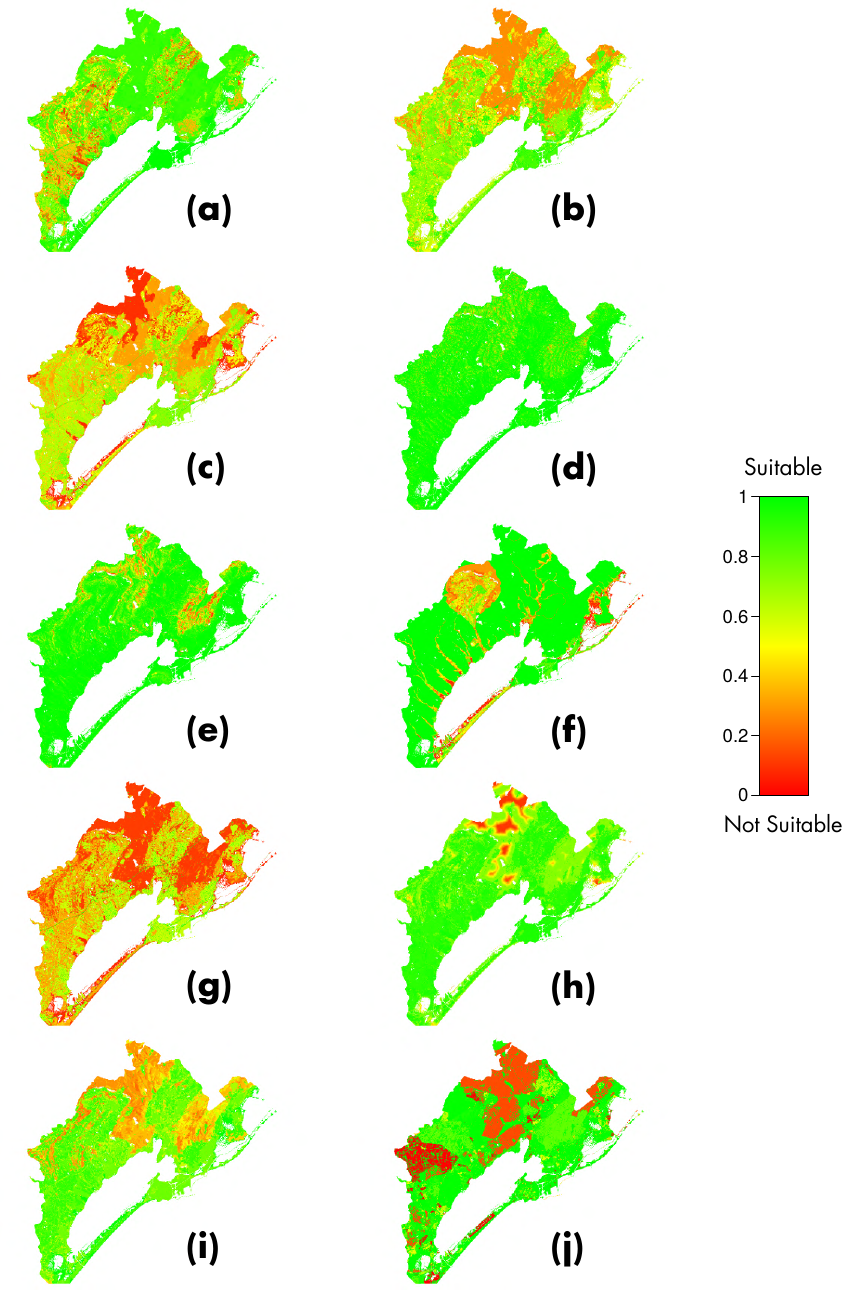}
		\caption{\sf \textbf{Map of the criteria.} The value are comprised between 0 and 1 assessing the suitability of each pixel to urban development inversely proportional to its capacity to provide the ES. (a) Cultivated crops. (b) Wild plants, algae, fungi and their outputs. (c) Maintaining nursery populations and habitats. (d) Maintenance of water quality. (e) Mass stabilization and control of erosion. (f) Flood Protection. (g) Aesthetic. (h) Physical and experiential interactions. (i) Wildfires. (j) Habitat patches importance. \label{FigS1}}
	\end{center}
\end{figure*}

All the criterion maps are available in Figure S\ref{FigS1}.

\newpage
\subsection*{Supplementary figures}

\begin{figure*}
	\begin{center}
		\includegraphics[width=12cm]{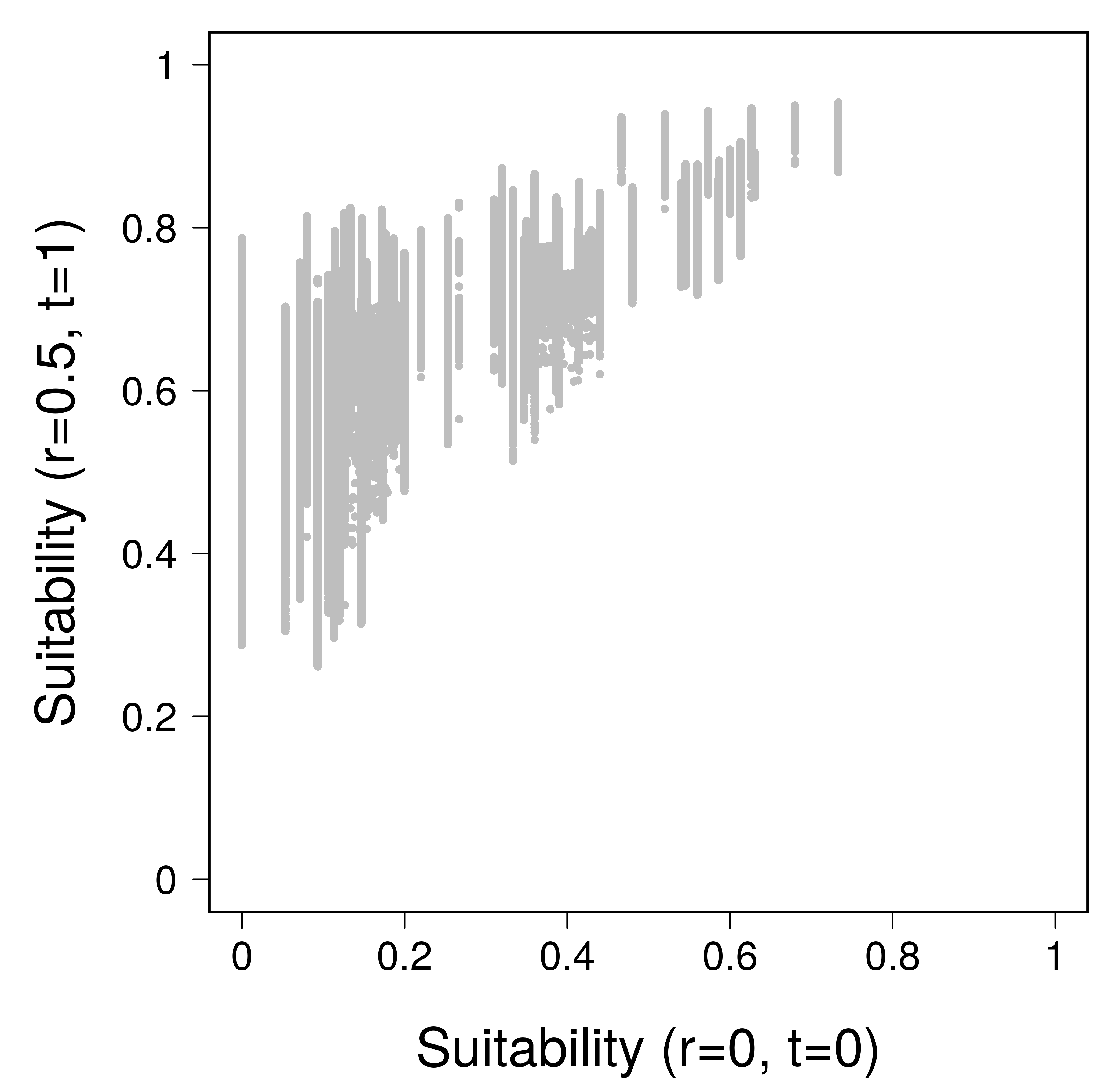}
		\caption{\sf \textbf{Comparison between suitability values obtained in the low risk with no trade-off configuration (r=0,t=0) and the intermediate risk with full trade-off configuration (r=0.5,t=1).} Each grey point represents a pixel. \label{FigS2}}
	\end{center}
\end{figure*}

\begin{figure*}
	\begin{center}
		\includegraphics[width=14cm]{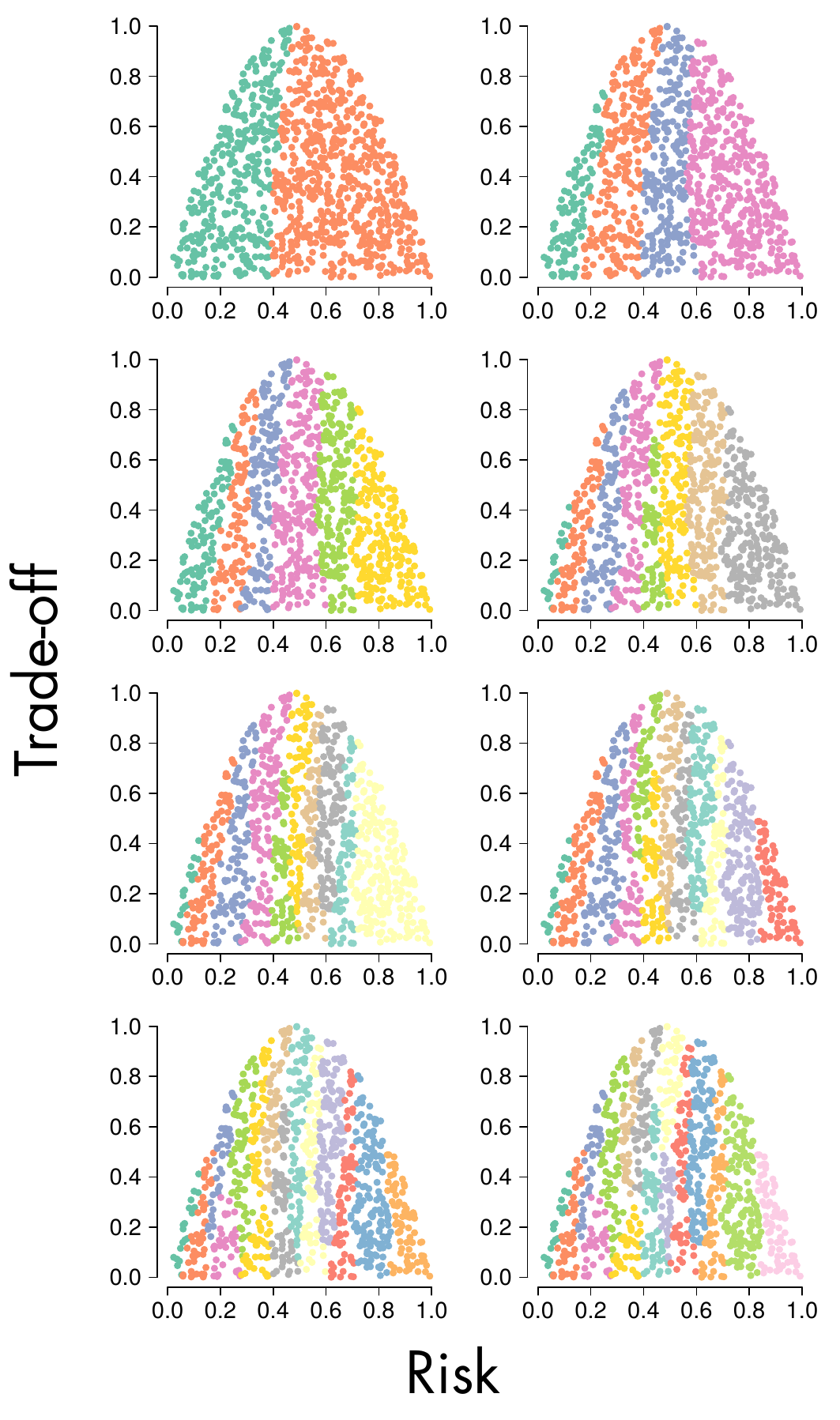}
		\caption{\sf \textbf{Clustering analysis.} Segmentation of the parabola decision-strategy space according to the number of clusters. \label{FigS3}}
	\end{center}
\end{figure*}

\begin{figure*}
	\begin{center}
		\includegraphics[width=16cm]{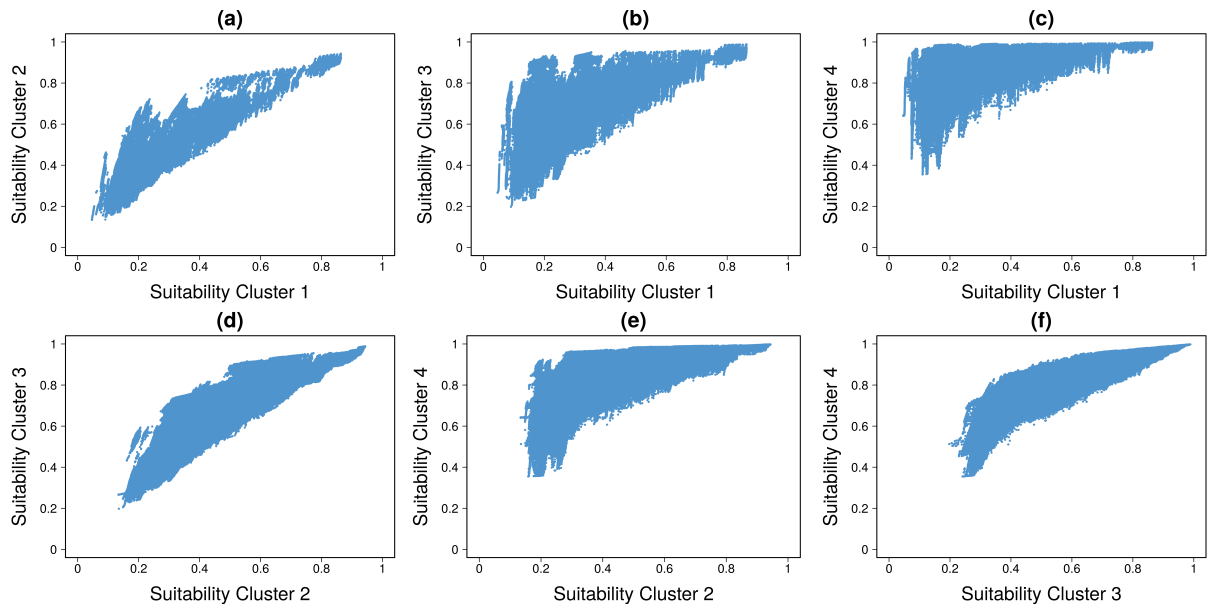}
		\caption{\sf \textbf{Comparison between suitability values obtained with the different clusters.} Each blue point represents a pixel. \label{FigS4}}
	\end{center}
\end{figure*}

\end{document}